\documentclass
[reprint,aps,amsmath,amssymb,longbibliography]{revtex4-2}%
\usepackage{amssymb}
\usepackage{graphicx}
\usepackage{amsmath}
\usepackage{amsfonts}%
\providecommand{\U}[1]{\protect\rule{.1in}{.1in}}
\begin{document}

\title{Partial persistence of memory in bubble breakup: incomplete universality
acquired by broken symmetry}
\author{Ikumi Yoshino and Ko Okumura\\Physics Department and Soft Matter Center, Ochanomizu University, 2-1-1
Ohtsuka, Bunkyo-ku, Tokyo 112-8610, Japan}
\date{\today}

\begin{abstract}
When a water drop falls from a faucet, the drop is created with the formation
of an axisymmetric constriction region, which thins down to breakup. Such
formation of a fluid drop has been extensively studied as a representative of
the singular dynamics widely observed in nature. The singular dynamics is
often self-similar, i.e., shapes at different times collapsing onto a master
curve after rescaling, and the self-similar dynamics has been categorized as
either \textit{universal or non-universal}: the master curve is
\textit{independent of or dependent on} the length scales that set the initial
boundary conditions, as if memory is \textit{erased or retained}. Here, we
focus on the post-breakup dynamics and confine the system to break the
axisymmetry, introducing three length scales, which leads to a third category
of incomplete universality, where memory is partially retained: the master
curve could be dependent on the smallest scale but independent of the other
two scales. Affecting of only the smallest length scale on the master curve
underscores the importance of scale separation for the emergence of
universality. The present study suggests a promising direction for the study
on the singular dynamics by exploring the symmetry.

\end{abstract}
\maketitle

Singular dynamics observed widely in nature from the gravitational collapse of
a star \cite{choptuik1993universality} to the formation of a drop in dripping
faucets \cite{1994ScienceNagelDropFallingFaucet} have attracted scientist in
many fields such as cosmology \cite{koike1995critical}, hydrodynamics, physics
and mathematics
\cite{Barenblatt1979,barenblatt2003scaling,eggers2015singularities}. One
important feature of the singular dynamics governed by partial differential
equations (PDEs) is \textit{self-similarity}, reported in various phenomena
including fluid-jet formation
\cite{2000NatureLathropFluidJetEruption,2002PRLCohenSelectiveWithdrawal} and
drop coalescence
\cite{YokotaPNAS2011,hernandez2012symmetric,kaneelil2022three}. The
self-similarity is expressed as $h(t,x)=h_{0}(t)\Gamma(x/x_{0}(t))$ for the
solution $h(t,x)$ of a PDE: although the shape defined by the plot $x$ vs
$h(t,x)$ for a given $t$ changes with $t$, the rescaled shape defined by the
plot $X=x/x_{0}(t)$ vs $Y=h(t,x)/h_{0}(t)$ at different times collapse onto a
master curve $Y=\Gamma(X)$.

The self-similar dynamics is said to be \textit{universal} or \textit{losing
memory}, when the master curve $\Gamma(X)$ is independent of the length scales
that set the initial boundary conditions (BCs). This scenario is expected from
the viewpoint of separation of scales: if the physics at small scales in
length and time becomes important near the singularity, the dynamics may
become independent from larger scales setting the BCs, as if memory (of the
initial condition) is lost. The opposite case, in which $\Gamma(X)$ is
dependent on the length scales, is said to be \textit{non-universal} or
\textit{retaining memory}. This scenario is expected when the governing
equation cannot define internal length scales.

One example of \textit{the universal self-similar dynamics losing memory} is
the breakup of a drop surrounded by less viscous fluid, where capillarity,
inertia, and viscosity all come into play \cite{1993PRLEggersPinchoff}: While
the rescaled shape at different times collapse onto a master curve, the master
curve is found to be independent of the single length scale that sets the BCs,
the radius of a tube to create the drop. However, the breakup of bubble
surrounded by more viscous fluid was shown to be \textit{a non-universal
self-similar dynamics retaining memory}, where the dynamics governed by the
Stokes equation describing the viscous-capillary balance
\cite{2003ScienceNagelMemoryDropBreakup}: The master curve does depend on the
tube radius, reflecting that the Stokes equation cannot define length scales.
Recently, this non-universal self-similar dynamics is found to exhibit
crossover to a new universal self-similar dynamics governed by a wetting rim
dynamics based on another viscous-capillary balance
\cite{pahlavan2019restoring}.

Here, we focus on the post-breakup dynamics and report a novel scenario in the
singular dynamics: \textit{the incomplete universality} where \textit{memory
is partially retained}. The key is a confinement, which breaks the axisymmetry
of the previous cases. As a result, three independent length scales are
introduced for the BCs. This is in contrast with the previous axisymmetric
cases, which allow only one length that sets the BCs, the tube radius. The
existence of three scales leads to a novel self-similar dynamics, in which the
master curve depends on the smallest length but not on the other scales: the
universality is incomplete with memory partially retained. This is also
expected from separation of scales if the length scale on which the collapse
to a master curve is observed is comparable to the smallest length scale
setting BCs but much smaller than the other scales.

We show an example of this new class of the singular dynamics by experiments
with a high quality. The novel scenario of the incomplete universality on the
memory in singular dynamics revealed in the present study opens a new avenue
for understanding singular dynamics widely observed in nature and underscores
the importance of exploration of the symmetry in singular dynamics.

\begin{figure}[ptb]
\centering
\includegraphics[width=0.45\textwidth]{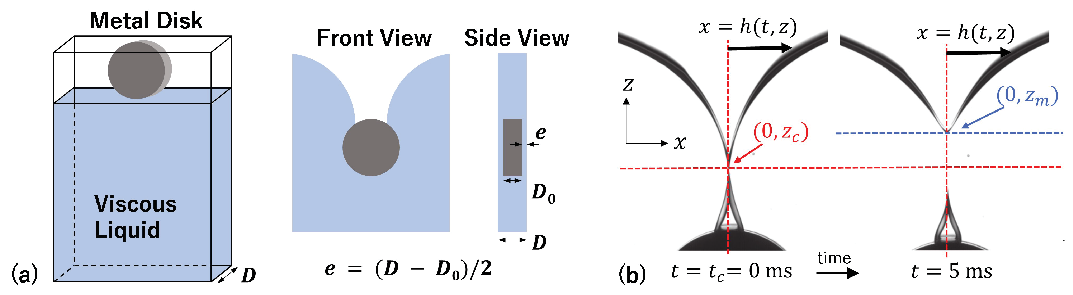}\caption{(a) Experimental
setup. A metal disk of thickness $D_{0}$ ($=2.0-3.5$ mm) and radius $R$
($10~$to $15$ mm) falls in the cell of thickness $D$ ($3$ to $4.5$ mm) filled
with a viscous liquid of kinematic viscosity $\nu$ (1 to 50 St). The disk
entrains air into the liquid, which finally detaches from the disk. The
difference between $D$ and $D_{0}$ defines the liquid film thickness $e$,
indicated in the side view from the right edge of the front view. (b)
Snapshots at breakup and after breakup illustrating the setting of axes for
$(e,D_{0},R,\nu)=(0.5,3,10,1)$ in mm or St. $z=z_{c}$ is set to the origin of
the $z$ coordinate.}%
\label{Fig1}%
\end{figure}


\textit{Experimental\label{S2}---} The setup and experimental procedure are
shown in Fig.~\ref{Fig1} (a). We regard the phenomenon as a confined bubble
breakup because, after air breaks up, the lower part detaches from the disk
and becomes a bubble. The cell width and height (typically 9 and 12 cm,
respectively) are much larger than the other length scales: the depth (or
thickness) of the cell $D$, the thickness $D_{0}$ and the radius $R$ of the
disk. Note that $D$ does not represent a diameter (but the cell depth), and
$D_{0}$ is comparable to and a bit smaller than $D$ with $e$ $=(D-D_{0})/2$
$<D_{0}$. We use polydimethylsiloxane (PDMS) for viscous liquid of kinematic
viscosity $\nu=\eta/\rho$. The density $\rho$ and the surface tension $\gamma$
are slightly depending on viscosity $\eta$ ($\rho\simeq$ $0.97$ g/cm$^{3}$ and
$\gamma\simeq20$ mN/m). The density $\rho_{s}$ of the metal disk is either is
7.7 g/cm$^{3}$ (stainless: SUS430) or 8.7 g/cm$^{3}$ (brass) with the density
difference $\Delta\rho=\rho_{s}-\rho$. The cell is fabricated with acrylic
plates of thickness 5 mm, using acrylic spacers whose thickness defines $D$.
To capture the dynamics, we analyzed the images with Image J and self-made
Python codes, which were obtained with a high-speed camera (FASTCAM Mini UX
100, Photron) with a lens (Micro NIKKOR 60 mm f2.8G ED, Nikon) at 1000 to 2000
frames per second (fps). The spacial resolution of the image is typically 50
pixels per mm or more.

We here summarize important points to realize high reproducibility. We need to
achieve (1) the thickness $e$ of the lubrication films on both surfaces of the
disk to be equal, i.e., $e=(D-D_{0})/2$, as indicated in Fig.~\ref{Fig1} (a),
(2) the initial velocity of the metal disk to be zero at the entry (initial
velocity affects the results, especially for the pre-breakup dynamics), and
(3) the contact angle of the surface of the disk to be zero (the contact angle
affects the results). Accordingly, (1) we set a gate with the gap $D_{0}$ at
the top of cell by placing two small plates of thickness $e$ on the inside
surfaces of the cell, (2) we fall the disk so that the bottom of the disk is
initially in contact with the air-liquid interface, and (3) we wipe the oil on
the surface of the disk after dipping to coat the surface by a thin layer of
the oil.

\begin{figure}[ptb]
\centering
\includegraphics[width=0.45\textwidth]{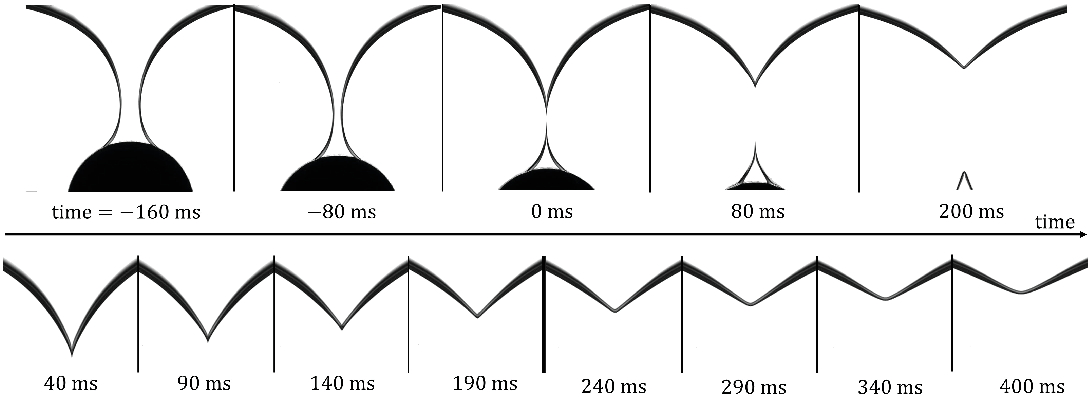}\caption{Snapshots of
entrainment of air by a disk into liquid, leading to breakup of a sheet of air
for $(R,D_{0},e,\nu)=(10,3,0.5,10)$ in mm or St. In the top panel, overall
time development is shown. In the bottom magnified snapshots, the sharp tip at
short times become rounded with time, with the thick vertical line in the
middle indicates the border. The time label 0 ms corresponds to $t=t_{c}$
defined in the text. }%
\label{Fig2}%
\end{figure}

\textit{Temporal change of the dynamics---} In Fig.~\ref{Fig2}, we show
typical snapshots before and after breakup in the present parameter range. A
thin air film is formed slightly before breakup, as confirmed by the side-view
snapshots obtained for a similar parameter set in our previous study on the
pre-breakup dynamics \cite{nakazato2018self}. In the post breakup of the
present focus, the tip is sharp when seen from the front as in the snapshots
at short times (just after the breakup), but becomes rounded with time, while
our focus below is mainly on the regime after the breakup but before the tip
becomes clearly rounded.

\textit{Shape of the air-liquid interface $h(z,t)$---} Figure \ref{Fig1} (b)
explains the definition of the shape and the setting of space-time coordinate.
The shape of air-liquid interfaces formed by air entrained by the disk are
seen as the dark line with a finite thickness. In the present study, the inner
edge of the right (or left) line is described by: $x=h(z,t)$ (or $-h(z,t)$).
For further details, see Appendix A1.

The space-time coordinate is determined as follows (see Fig.~\ref{Fig1} (b)).
Before breakup the function $h(z,t)$ possesses a minimum with respect to $z$,
which we call "the constriction point," at which $(x,z)=(h_{m}(t),z_{m}(t))$,
i.e., $h_{m}(t)=h(z_{m}(t),t)$. At $t=t_{c}$, topology changes: The fluid
breaks into two chunks at the constriction point $(h_{m}(t_{c}),z_{m}%
(t_{c}))=(0,z_{c})$, where $z=z_{c}$ will be set to the origin of the $z$
coordinate. After $t=t_{c}$, the constriction point thus disappears and the
dynamics of the interface of our focus is characterized by the tip point
$(0,z_{m}-z_{c})=(0,z_{m})$. Experimentally, our time label $0$ ms could
deviate from the true $t=t_{c}$ at most 1 ms, which is set by frames per
second in capturing images. See further details for Appendix A2.

The way of selecting the edge and the determination of time label $0$ ms,
detailed in Appendix A1 and A2, are technically important for data collapse
presented below. Slight changes could deteriorate the quality of the collapse.

\begin{figure}[ptb]
\centering
\includegraphics[width=0.45\textwidth]{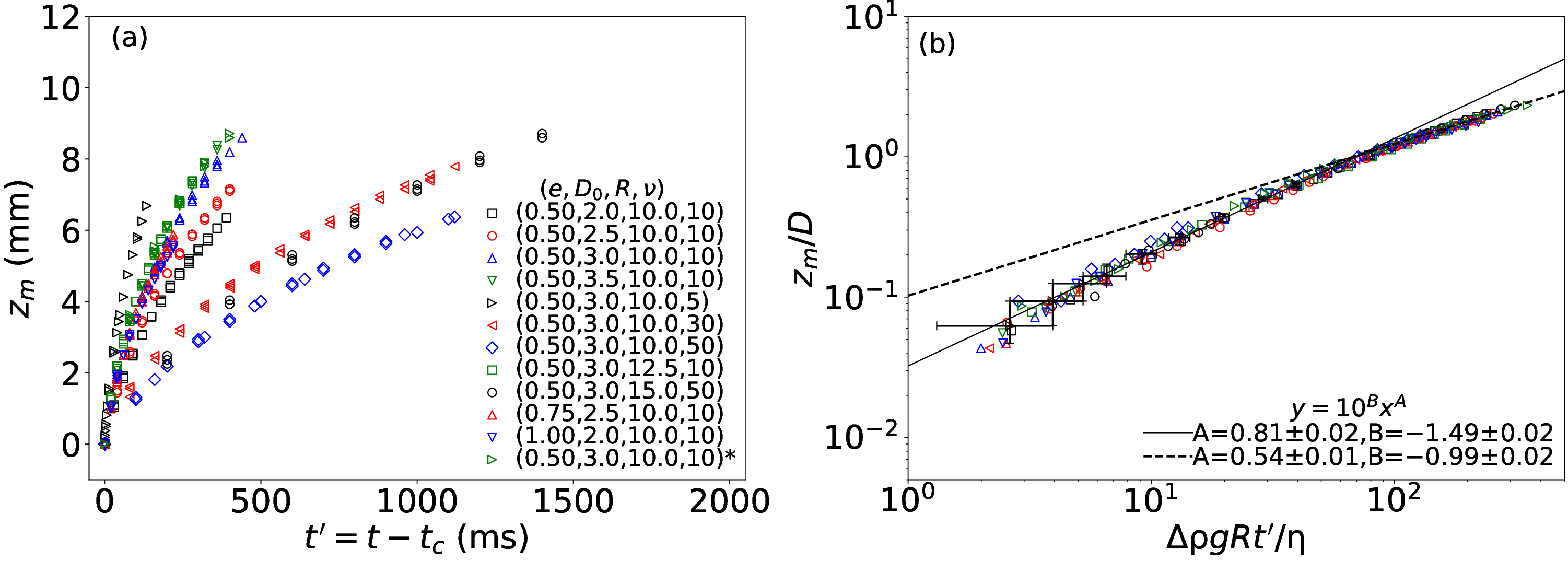}\caption{(a) $z_{m}$ vs.
$t^{\prime}=t-t_{c}$, where $t_{c}$ is the critical time defined in the text
for various parameters $(e,D_{0},R,\nu)$ in mm or St. The data marked with a
star (*) in the legend are those obtained for a different $\Delta\rho$ by
using a brass disk instead of a disk of stainless steel. (b) Distinct data
collapse by Eq.~(\ref{eq1c}). All the data in (a) are plotted on rescaled
axes, based on Eq.~(\ref{eq1c}), demonstrating a clear data collapse with a
quasi scaling-crossover between the regimes reasonably well characterized by
the exponent $\Delta^{\prime}\simeq0.8$ and 1/2, as indicated by the results
of fitting where $y=z_{m}(t)/D$ and $x=\Delta\rho gRt^{\prime}/\eta$.}%
\label{Fig3}%
\end{figure}

\textit{Dynamics of characteristic length scale---} In Fig.~\ref{Fig3} (a), we
present the relation between $z_{m}$ and $t$ for various conditions, in which
we can confirm an excellent reproducibility of the present measurement. For
example, if we closely examine the data shown by blue diamond, we see several
overlapping data points, which are generally obtained on different days.

In Fig.~\ref{Fig3} (b), in which we show error bars for a set of data, we can
confirm all the data in (a), which include the data for different $e$, $D_{0}%
$, $R$, $\eta$, and $\Delta\rho$, can be well described by the following
relation between dimensionless quantities:
\begin{equation}
z_{m}(t)/D=f\left(  \Delta\rho gRt^{\prime}/\eta\right)  \equiv f(t^{\prime
}/\tau) \label{eq1c}%
\end{equation}
with $\tau=\eta/\Delta\rho gR$, where a time label $t^{\prime}$ (which is
positive at times after $t=t_{c}$) is defined as $t^{\prime}=t-t_{c}$ (The
dependence on $\Delta\rho$ is more explicitly shown in Fig.~\ref{Fig6} (b3)
below). We can further confirm in Fig.~\ref{Fig3} (b) that the slope
$\Delta^{\prime}$ of the $z_{m}$-$t^{\prime}$ relation on log-log scales seem
to exhibit a crossover from Regime I to Regime II: $\widetilde{z}%
_{m}=\widetilde{t}^{\Delta^{\prime}}$with $\widetilde{z}_{m}=z_{m}/D$ and
$\widetilde{t}\simeq t^{\prime}/\tau$, where $\Delta^{\prime}\simeq0.8$ in
Regime I and $\Delta^{\prime}\simeq1/2$ in Regime II (results of fitting are
shown in the plot), although the range of regimes are limited (especially in
Regime II).

\begin{figure}[ptb]
\includegraphics[width=0.45\textwidth]{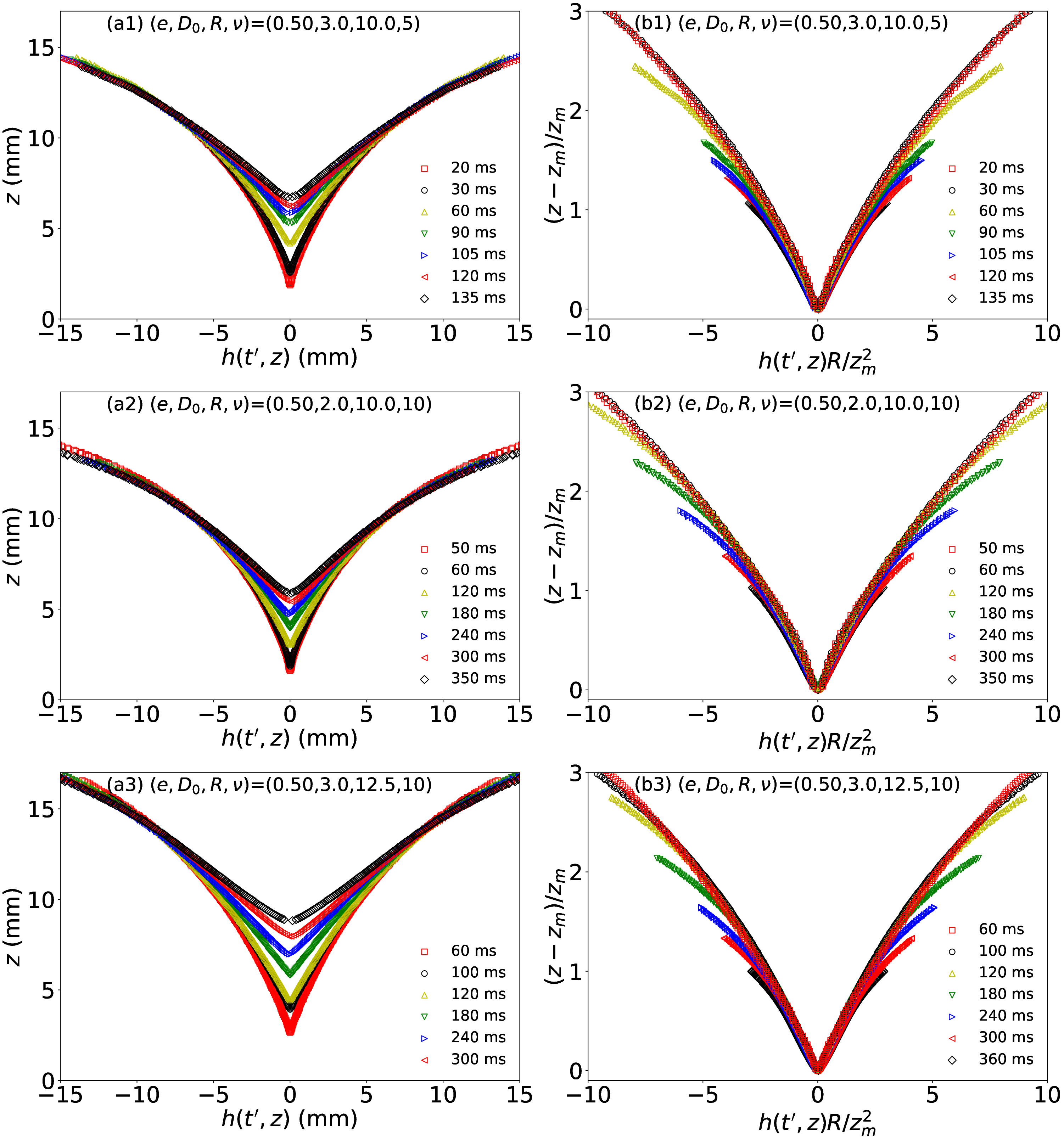}\caption{(a1) to (a3) Temporal
change of the interface $h(z,t)$ at $e=0.5$ mm for the parameter sets
$(D_{0},R,\nu)=(3,10.0,5)$, $(1.0,10.0,10)$ and $(2.0,12.5,10)$ in mm or St.
(b1) to (b3) The space-time dependent collapse of the shape by
Eq.~(\ref{eq18a}). The collapse near the tip persists for long time, while
that away from the tip is observed only for short time.}%
\label{Fig4}%
\end{figure}

\textit{Self-similarity in the interface shape dynamics---} In Fig.~\ref{Fig4}
(a1) to (a3), we show temporal changes of the interfacial shape after breakup
for $e=0.5$ mm but with different $\eta$, $D_{0}$, and $R$. As seen in
Fig.~\ref{Fig4} (b1) to (b3), interface shapes after rescaling both axes by
$h(z,t)R/z_{m}^{2}$ and $(z-z_{m})/z_{m}$ are clearly collapsed onto a master
curve, especially near the tip where $(z-z_{m})/z_{m}\,\lesssim1$. The
collapse shown in (b) implies the following scaling form:%
\begin{equation}
h(z,t)=\frac{z_{m}^{2}}{R}\Gamma\left(  \frac{z-z_{m}}{z_{m}}\right)
\label{eq18a}%
\end{equation}

\begin{figure}[ptb]
\includegraphics[width=0.45\textwidth]{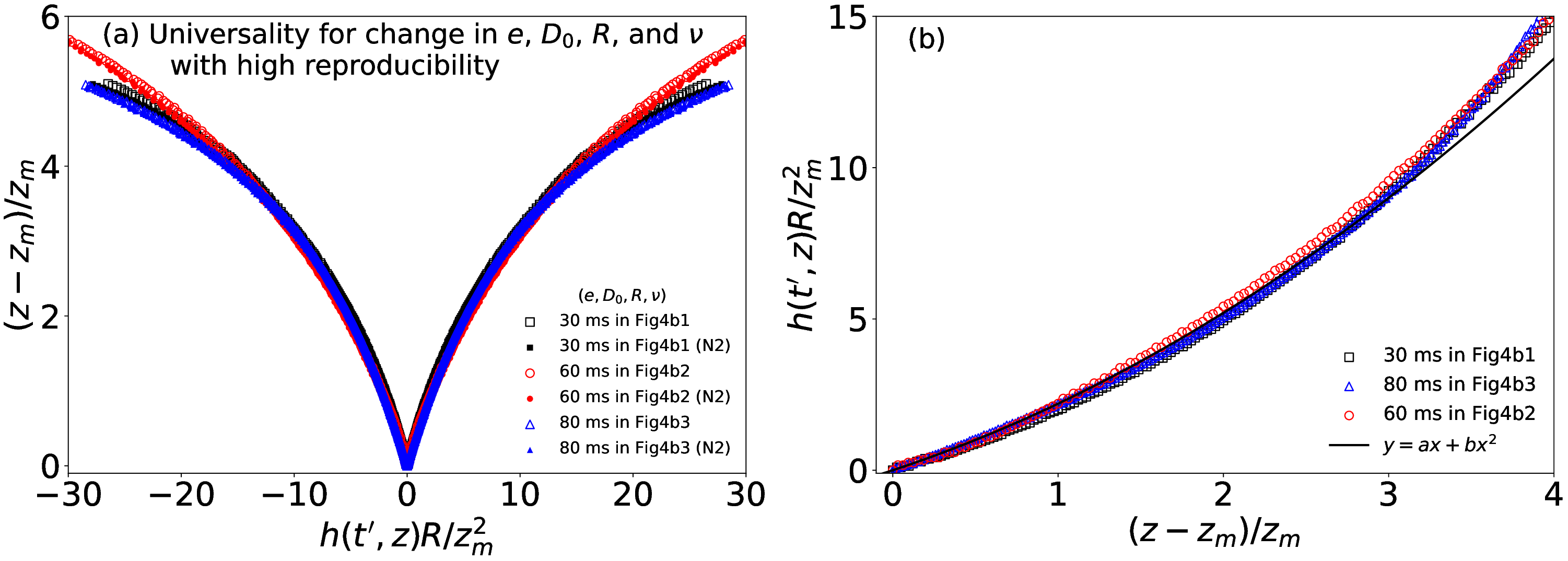}\caption{(a) The master curves
universal (and highly reproducible) for change in $D_{0}$, $R$, and $\nu
=\eta/\rho$ at $e=0.5$ mm (but non-universal for change in $e$ and $\Delta
\rho$, as shown below in Fig.~\ref{Fig6} (a3)). Shapes at earlier times in
Fig.~\ref{Fig4} are compared. The data marked as (N2) are basically obtained
on a different day. (b) Right branches in (a) are shown with the axes
interchanged, together with the results of fitting, where $y=h(z,t)R/z_{m}%
^{2}$ and $x=(z-z_{m})/z_{m}$.}%
\label{Fig5}%
\end{figure}

As shown in Fig.~\ref{Fig5} (a), the master curve near $t=t_{c}$ collapses
well (and thus is reproducible) beyond the linear region near the tip up
roughly to $(z-z_{m})/z_{m}=3$ even if parameters $D_{0}$, $R$, and $\eta$ are
changed in a limited range with $e$ fixed to 0.5 mm. The universal master
curve seems to be linear near the tip but quadratic away from the tip. In
fact, we can show the master curve is well described by $\Gamma(x)=ax+bx^{2}$
with $a=1.8\pm0.2$ and $b=0.4\pm0.05$ as shown in Fig.~\ref{Fig5} (b) by a
two-step fitting: We first determine $a$ by fitting to the linear region near
the tip (on log-log scales) and then determine $b$ by fitting to the quadratic
region after subtracting the linear component. Note that the confirmation of
the collapse by Eq.~(\ref{eq18a}) becomes very difficult for the shape very
close to $t=t_{c}$, such as those at $t$ with $t-t_{c}\lesssim$ 10 ms, which
is the order of the time resolution of our experiment limited by frames per
second in capturing images. This is because $z_{m}$ becomes very small and
approaches the spacial resolution of our experiment, meaning that the error in
estimating $z_{m}$ becomes too large.

The collapse observed for the post-breakup dynamics ($t>t_{c}$) in
Fig.~\ref{Fig4} is regarded as space-time dependent if we notice that, by the
present rescaling $h(z,t)R/z_{m}^{2}$ and $(z-z_{m})/z_{m}$, shapes at times
closer to $t=t_{c}$ are more strongly magnified, since $z_{m}$ becomes smaller
as $t\rightarrow t_{c}$: (1) Near the tip ($z=z_{m}$), the master curve is
linear and the shape collapse to it persists for times rather away from the
breakup ($t=t_{c}$). (2) Away from the tip ($z>z_{m}$), it is parabolic, but
the collapse is limited only near the breakup time ($t\simeq t_{c}$). The
space-time dependence is summarized as%

\begin{equation}
\Gamma(X)\simeq\left\{
\begin{array}
[c]{ccc}%
X & (X<1) & \text{ for long time after }t=t_{c}\\
X^{2} & (X>1) & \text{ only for short time after }t_{c}%
\end{array}
\right.  \label{eq18b}%
\end{equation}

In other words, in Figs. \ref{Fig4} (and \ref{Fig6} below), the master curve
is confirmed in the spacial region where the rescaled curves at the earliest
time and that at the second earliest collapse: The spacial region of the
master curve confirmed in this way is up roughly to $(z-z_{m})/z_{m}=3$.

\begin{figure}[ptb]
\includegraphics[width=0.45\textwidth]{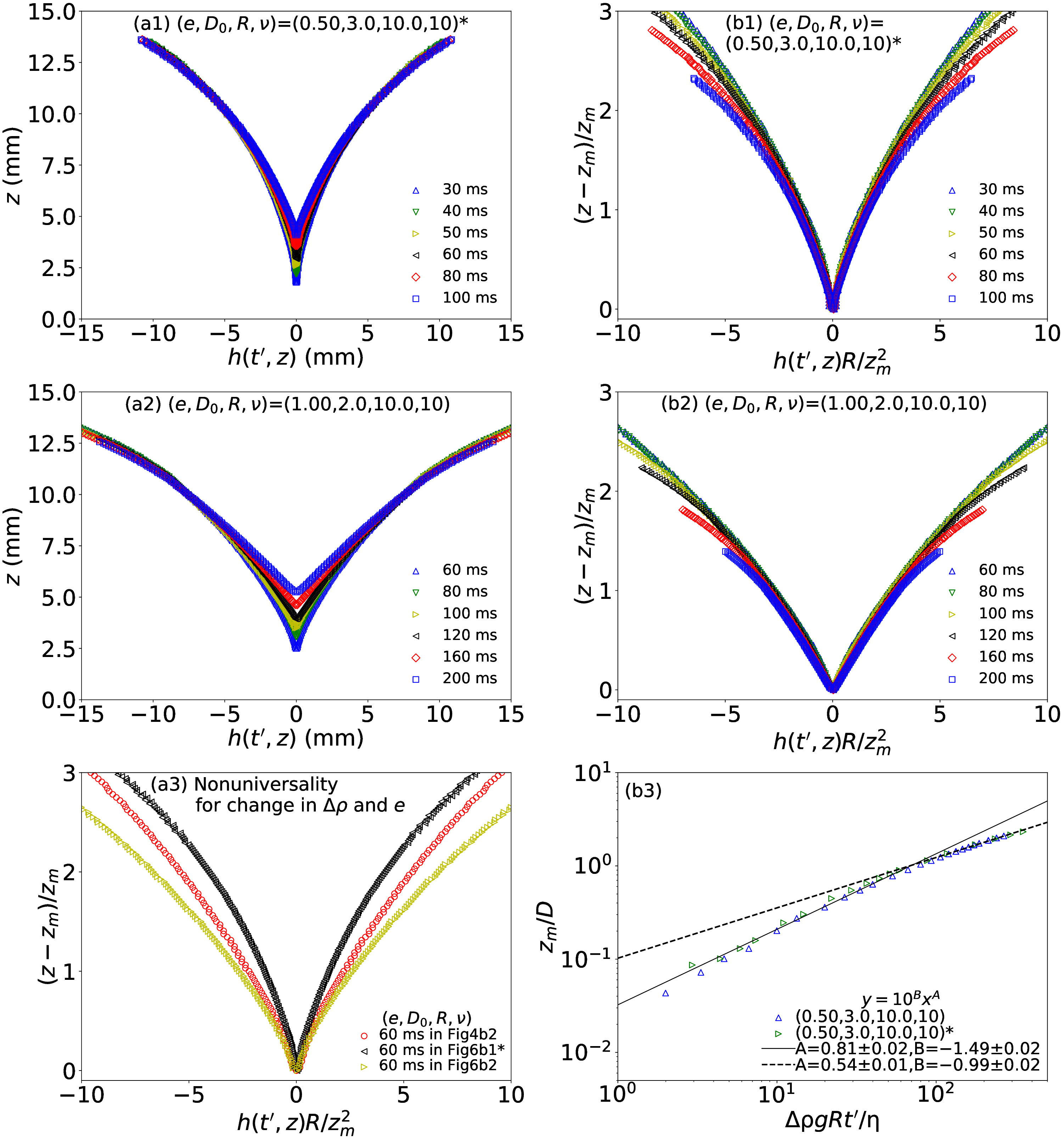}\caption{(a1) and (a2)
Temporal change of the interface $h(z,t)$ at $\Delta\rho$ and $e$ different
from those in Fig.~\ref{Fig4} or Fig.~\ref{Fig5}: $(e,D_{0},R,\nu
)=(0.5,3,10,10)$ in mm or St using a brass disk and $(1.0,2.0,10,10)$ using a
stainless disk. The star (*) in (a1) [as well as in (b1), (a3) and (b3) below]
indicates the data obtained for different $\Delta\rho$ as in Fig.~\ref{Fig3}.
(b1) to (b2) The space-time dependent collapse of the shape by
Eq.~(\ref{eq18a}). (a3) The master curves non-universal for change in $e$ and
$\Delta\rho$. Shapes in (b1) and (b2) are compared with a shape in
Fig.~\ref{Fig4} (b2) at an earlier time. (b3) $z_{m}$ vs. $t^{\prime}=t-t_{c}%
$. We extract from Fig.~\ref{Fig3} (b) the data obtained for two different
$\Delta\rho$'s (using the stainless and brass disks) at $(e,D_{0}%
,R,\nu)=(0.5,3,10,10)$ in mm or St with the fitting lines in Fig.~\ref{Fig3}
(b), to clearly show the $\Delta\rho$ dependence. The results of the fitting
are also shown with $y=z_{m}(t)/D$ and $x=\Delta\rho gRt^{\prime}/\eta$.}%
\label{Fig6}%
\end{figure}

The space-time dependent behavior of the master curve in Eq.~(\ref{eq18b}) is
preserved even if we change $e$ and $\Delta\rho$ in a certain range, as shown
in Fig.~\ref{Fig6} (a1) to (b2). However, the master curve defined by shapes
at earlier times in the range up roughly to $(z-z_{m})/z_{m}=3$ clearly
depends on $e$ and $\Delta\rho$ as demonstrated in Fig.~\ref{Fig6} (a3), where
we compare the shape at one early time with those obtained for a different $e$
and a different $\Delta\rho$. We have also confirmed that the master curve at
$e=0.75$ mm, for example, collapses onto neither the master curve at $e=0.5$
mm nor that at $e=1.0$ mm for a fixed $\Delta\rho$.


\textit{The incomplete universality: partial persistence of memory---} The
independence of the master curve from the length scales $D_{0}$ and $R$ shown
in Fig.~\ref{Fig5} (a) and dependence on $e$ shown in Fig.~\ref{Fig6} (a3)
reveal a new category of the incomplete universality for the memory of
singular dynamics discussed in Introduction. This is because these lengths are
the scales that set the initial boundary conditions of the present problem. In
the present case, the master curve loses the memory on the scales $D_{0}$ and
$R$, but retains the memory of the smallest scale $e$, which is quite natural
from the viewpoint of scale separation, with remaining dependence on the
material parameter on $\Delta\rho$ (and independence from $\eta$).

As a matter of fact, Fig.~\ref{Fig6} (a3) shows that the two curves obtained
for different $e$'s but the same $\Delta\rho$ become indistinguishable near
the tip (although they seem distinguishable from the other obtained for a
different $\Delta\rho$). This observation is in accord with our result of the
two-step fitting: $(a,b)=(0.9\pm0.1,0.6\pm0.05)$ for Fig. \ref{Fig6} (b1) and
$(1.9\pm0.2,0.7\pm0.05)$ for Fig. \ref{Fig6} (b2): $a=1.9$ is
indistinguishable from $a=1.8$ (although $a=0.9$ is distinguishable). This
observation that near the tip universality tends to recover is also expected
from separation of scales: very close to the tip, the characteristic length
for the collapse tends to become small and thus become well separated even
from the smallest length scale $e$.

\textit{The pre- and post-breakup dynamics---} In the case of the
viscous-capillary breakup of a bubble under no geometrical constraint, where
the axisymmetric liquid-air interface is described by $r=h(z,t)$ in the
cylindrical coordinate $(r,\theta,z)$, the self-similar structure in the pre-
and post-breakup can be expressed in the same form (see, e.g., Sec.~3.5.1 of
\cite{eggers2015singularities}): $h(z,t)=|\widetilde{t}^{\prime}%
|\Gamma(z/|\widetilde{t}^{\prime}|^{2})$ with a dimensionless time
$|\widetilde{t}^{\prime}|\sim\left\vert t-t_{c}\right\vert $. This is in
contrast with the confined case. The scaling structure for the pre-breakup,
reported as $h(z,t)=2|\widetilde{t}^{\prime}|\Gamma(z/|\widetilde{t}^{\prime
}|)$ in \cite{nakazato2018self}, is significantly different from that for the
post-breakup, $h(z,t)\simeq|\widetilde{t}^{\prime}|^{2\Delta^{\prime}}%
\widehat{\Gamma}(z/|\widetilde{t}^{\prime}|^{\Delta^{\prime}})$, as seen from
Eq.~(\ref{eq18a}). This difference in structure may be related the elongation
of the constriction region near breakup. Because of the elongation, the
critical time leading to a good collapse in the post-breakup becomes different
from that in the pre-breakup.

\textit{Physics at the level of dimensional analysis---} We may regard the
present problem as finding a solution for Navier-Stokes equation for a viscous
liquid, neglecting the role of air, giving initial BCs. It would be easier to
give the initial BCs at the time $t=t_{c}$ rather than the time of the entry
of the disk because after $t=t_{c}$ we can forget about the disk motion by
giving the velocity of the tip of air at $t=t_{c}$. This velocity is given as
$v_{G}\sim\Delta\rho gD^{2}/\eta$ by balancing gravitational and viscous
energy (per time) for the disk, $\Delta\rho gR^{2}Dv_{G}\sim\eta(v_{G}%
/D)^{2}R^{2}D$ (for $D\simeq D_{0}$), in the previous study on the pre-breakup
dynamics \cite{nakazato2018self}. In addition, from Eqs.~(\ref{eq18a}) and
(\ref{eq1c}) with the dependence on $e$ of the master curve, we expect the BCs
involves $\Delta\rho,\eta,g,R,e,$and $D$ but not on $\gamma$. Thus, we expect
$h=f(t^{\prime},z,\Delta\rho,\eta,g,R,e,D)$. Here, we have 9 dimensional
variables, of which only 6 are independent, since the dimension of the unit of
all variables can be derived from the three fundamental units, kg, m, and s.

From the Buckingham $\pi$ theorem (see, e.g., Appendix C of
\cite{eggers2015singularities}), we expect a relation $\pi_{0}=\Xi(\pi_{1}%
,\pi_{2},\ldots,\pi_{5})$, where $\pi_{i}$'s are 6 independent dimensional
variables and $\Xi$ is a dimensionless function. We select these dimensionless
variables as follows. A natural characteristic scale in the $z$ direction is
$z_{m}$, from which we define $\pi_{1}=z/z_{m}-1$. A natural unit $h^{\ast}$
(in the $x$ direction) for $h$ can be introduced through a curvature relation,
$1/R\sim h^{\ast}/z_{m}^{2}$, from which we set $\pi_{0}=h/h^{\ast}$. To
select the remaining 4 independent variables, we focus on 4 length scales:
$l=\eta/(\Delta\rho gt^{\prime})$, $R$, $e$, and $D$, which are normalized by
$z_{m}$ to determine 4 dimensionless variables, $\pi_{2},\ldots,\pi_{5}$. In
this way, to be consistent with our experiment, we may arrive at
$h=\frac{z_{m}^{2}}{R}\Xi((z-z_{m})/z_{m},l/z_{m},R/z_{m},D/z_{m},e/z_{m})$.
Here, we may expect that near the breakup point where $z_{m}\sim(t^{\prime
})^{\Delta^{\prime}}$ with $0<\Delta^{\prime}<1$ is small so that the
right-hand side of the equation becomes independent of the second to the
fourth dimensionless variables (with $l/z_{m}\rightarrow\infty$ and
$R/z_{m},D/z_{m}\rightarrow0$ but with a finite $e/z_{m}$).

In this way, the dimensional analysis provides a natural understanding of the
scaling structure in Eq.~(\ref{eq18a}) with the function $\Gamma$ dependent on
$e$ but not on $R$ and $D$, based on which structure we proposed the novel
scenario of the incomplete universality. Although the present analysis cannot
determine the scaling exponents (such cases are known as the self-similarity
of the second kind \cite{barenblatt2003scaling}), a renormalization group
analysis recently developed for the non-confined bubble breakup
\cite{Okumura2025RG} will be promising to determine the exponents and to
elucidate the origin of the scaling crossover in the present study.

\textit{Importance of exploring symmetry---} We stress here that it is
becoming clear that the present experimental system that can break the
axisymmetry contains extremely rich physics. In the previous studies
\cite{nakazato2018self,nakazato2022air,Ii}, three distinct regimes have been
found by changing the range of confining length scales: the sheet-forming
regime with breakup, the corn-forming regime with and without breakup. In the
first and second, a sheet and a corn of air (which are both non-axisymmetric)
are respectively formed at the constriction point leading to breakup, while,
in the third, a corn detaches from the disk without appearance of the
constriction point. In the third, in particular, analogy with critical
phenomena is deeply explored to find the exponents are dependent on a length
scale, which is a novel feature not found in the first and second regimes. In
all of the three studies \cite{nakazato2018self,nakazato2022air,Ii}, we focus
on the pre-detachment dynamics where detachment includes breakup. In the
present study, we focus on the post-breakup for the first time, while,
experimentally, the post-breakup dynamics has not been explored even in the
non-confined case. As a result, we found a self-similar structure
significantly different from those in pre-breakup dynamics as seen above and
successfully obtained the data strongly supports the novel scenario of the
incomplete universality.

The remarkable physical richness of the present experimental system was
uncovered by breaking of symmetry, which reminds the importance of symmetry
and dimensionality in critical phenomena \cite{Cardy,Goldenfeld}: a myriad of
universality classes have been found by exploring symmetry and dimensionality,
which has propelled the developments of modern physics ranging from soft and
hard condensed matter, non-equilibrium systems to active matter
\cite{de1979scaling,livi2017nonequilibrium,altland2023condensed,tailleur2022active}%
. Given this, the present study suggests a promising direction for the study
on the singular dynamics: \textit{exploring the symmetry in confined
geometries with drawing an analogy with critical phenomena}. This direction is
all the more promising if we remind that the singular dynamics has been widely
observed and confinement is involved in many cases of natural phenomena and
industrial processes ranging from geology and petroleum industry to
microfluidics with applications such as in medicine and biochemistry
\cite{parmigiani2016bubble,StoneStroockAjdari2004,HeleShawPetroleum2010,anna2016}%
.

\textit{Acknowledgments---} This work was supported by JSPS KAKENHI Grant
Number JP19H01859 and JP24K00596.


\section*{Appendix}

\textit{A1: Definitions of the shape function $h$---} The inner edge by which
we define the shape corresponds to the contact line on the front surface of
the back cell wall, whose surface is totally wetting (thin layer of oil exists
ahead of "the contact line"). The relation $x=h(z,t)$ represents the shape
outlined by the contact line on the back cell plate.

We here explain why the inner edge of the dark line corresponds to the contact
line. In fact, the shape of the liquid-air interface is three dimensional: it
should be a function of $y$: $x=\widetilde{h}(t,z;y)$ with $h(z,t)=\widetilde
{h}(t,z;y=D)$, where $y=0$ and $y=D$ respectively correspond to the back
surface of the front cell plate and the front surface of the back plate. This
implies that the curvature $\partial^{2}\widetilde{h}(t,x,z;y)/\partial y^{2}$
is negative (the surface of right branch is convex seen from the right liquid
side), which goes to zero towards the tip: the outer edge corresponds
$x=\widetilde{h}(t,z;D/2)$ where $y=D/2$ corresponds to the middle plane
between the inner surfaces of cell plates. This is because of continuity of
surface, which is horizontal at places far away from tip, where the surface
should be concave seen from the upper air phase. In addition, the contact line
at $y=D$ is further away from the viewpoint compared with the line at $y=0$,
when seen from the front.

\textit{A2: Setting of space-time coordinates---} The critical time $t_{c}$
used in the analysis was determined as follows. Towards breakup, the
constriction region starts to form a thin thread (when seen from the front),
which finally pinches off, by which moment, we define $t=t_{c}$. Precisely
speaking, the pinch-off is judged not by the inner edge but the outer edge as
indicated in Fig.~1 (b) left (the inner edge breaks up before the outer edge),
but the positions $z_{c}$ and $z_{m}$ are determined by the inner edge (In
Fig.1 (b), these positions are actually slightly above the dashed horizontal
lines by the amount of the thickness of the line representing the interface).
In addition, we set the time label $0$ ms as the snapshot just before the
pinch off, using snapshots obtained at 1000 frames per second, which means
there could be a difference at most 1 ms between our time label 0 ms and the
actual critical time $t=t_{c}$. Because of this possible error in determining
the origin of time, we avoid using snapshots too close to $t=t_{c}$ and use
only those at $t\gtrsim10$ ms.

\end{document}